%% file: main.tex
\documentclass[%
reprint,
superscriptaddress,
showpacs,
preprintnumbers,
nofootinbib,
amsmath,
amssymb,
amsfonts,
aps,
prd,
floatfix,
showkeys
]{revtex4-1}

\RequirePackage{atbegshi}

\input{Header}

\begin{document}

\title{
Quantum computational resources for lattice QCD in the strong-coupling limit
}

\author{Michael Fromm}
 \email{mfromm@itp.uni-frankfurt.de}
 \affiliation{
Institut f\"{u}r Theoretische Physik, Goethe-Universit\"{a}t Frankfurt\\
Max-von-Laue-Str.\ 1, 60438 Frankfurt am Main, Germany
}

\author{Lucas Katschke}
\email{katschke@itp.uni-frankfurt.de}
\affiliation{
Institut f\"{u}r Theoretische Physik, Goethe-Universit\"{a}t Frankfurt\\
Max-von-Laue-Str.\ 1, 60438 Frankfurt am Main, Germany
}

\author{Owe Philipsen}
 \email{philipsen@itp.uni-frankfurt.de}
 \affiliation{
Institut f\"{u}r Theoretische Physik, Goethe-Universit\"{a}t Frankfurt\\
Max-von-Laue-Str.\ 1, 60438 Frankfurt am Main, Germany
}
 \affiliation{
John von Neumann Institute for Computing (NIC) at GSI\\
Planckstr. 1, 64291 Darmstadt, Germany
}

\author{Wolfgang Unger}
 \email{wunger@physik.uni-bielefeld.de}
 \affiliation{
Fakult\"{a}t f\"{u}r Physik, Bielefeld University\\
33615 Bielefeld, Germany
}

\begin{abstract}
We consider the strong coupling limit of lattice QCD with massless staggered quarks and study the resource requirements for quantum simulating the theory in its Hamiltonian formulation. The bosonic Hilbert space of the color-singlet degrees of freedom grows quickly with the number of quark flavors, making it a suitable testing ground for resource considerations across different platforms. In particular, in addition to the standard model of computation with qubits, we consider mapping the theory to qudits $(d>2)$ and qumodes, as used on trapped-ion systems and photonic devices, respectively.
\end{abstract}


\keywords{Hamiltonian lattice gauge theory, Quantum computing}
\maketitle

\input{Intro}
\input{Background}
\input{Results}

\input{Conclusion}

\acknowledgments
The work is supported by the Deutsche Forschungsgemeinschaft (DFG, German Research Foundation) through the grant CRC-TR 211 ``Strong-interaction matter under extreme conditions''~--~project number 315477589~--~TRR 211 and by the State of Hesse within the Research Cluster ELEMENTS (Project ID 500/10.006). M.F. acknowledges support by the Munich Institute for Astro-, Particle and BioPhysics (MIAPbP) which is funded by the Deutsche Forschungsgemeinschaft (DFG, German Research Foundation) under Germany's Excellence Strategy – EXC-2094 – 390783311.

\bibliography{pub.bib}

\onecolumngrid

\appendix
\input{Appendix}

\end{document}

%% file: Header.tex
\usepackage[utf8]{inputenc}
\usepackage[T1]{fontenc}
\usepackage[ngerman,english]{babel}
\usepackage[matha]{mathabx} 
\usepackage[dvipsnames, svgnames, table]{xcolor}
\usepackage[braket, qm]{qcircuit}
\usepackage{bbm}
\usepackage{%
    graphicx,
    siunitx,
    multirow,
    booktabs,
    bbold,    
    pifont,   
    capt-of,
    pgfplotstable,
    todonotes,
}
\usepackage[version=4]{mhchem}
\usepackage[flushleft]{threeparttable}
\usepackage{subfigure}     
\usepackage[colorlinks]{hyperref}
\usepackage{makecell}
\usepackage{braket}
\hypersetup{
    colorlinks=true,
    citecolor=DarkRed,
    linkcolor=black,
    urlcolor=black,
    anchorcolor=black,
    linktocpage
}
\usepackage{outlines}
\usepackage{etoolbox} 
\usepackage{cleveref} 
\makeatletter
\appto{\appendix}{%
  \@ifstar{\def\theequation@prefix{A.}}%
          {}%
}
\makeatother
\crefname{figure}{Figure}{Figures}
\crefname{table}{Table}{Tables}
\crefname{equation}{Eq.}{Eqs.}
\crefname{section}{Section}{Sections}
\crefformat{appendix}{the #2Appendix#1#3} 
\graphicspath{{./figures/}}

\DeclareMathOperator{\Tr}{Tr}

\newcommand{\Nc}{N_\text{c}}
\newcommand{\Nf}{N_\text{f}}
\newcommand{\clqcd}{CL\kern-.25em\textsuperscript{2}QCD}
\newcommand{\beq} {\begin{eqnarray}}
\newcommand{\eeq} {\end{eqnarray}}

\newcommand{\eref}[1]{Eq.~(\ref{#1})}

\newcommand{\calH}{\mathcal{H}}
\newcommand{\calN}{\mathcal{N}}
\newcommand{\calZ}{\mathcal{Z}}
\newcommand{\bareT}{aT}
\newcommand{\bareMu}{a\mu_B}
\newcommand{\bareI}{a\mu_I}

\newcommand{\meson}{\mathfrak{m}}
\newcommand{\baryon}{\mathfrak{b}}
\newcommand{\hadron}{\mathfrak{h}}
\newcommand{\spin}{\mathfrak{s}}
\newcommand{\Hil}{{\mathbbm{H}_\hadron}}

\newcommand{\lr}[1]{\left(#1\right)}

%% file: Intro.tex
\section{Introduction}
It was a key observation that universal quantum computers offer the potential to simulate aspects of quantum systems efficiently that are intractable by classical machines~\cite{Lloyd1996}, provided the system's interaction remains local. As many of these are relevant for High Energy Physics, it is not surprising that quantum computation quickly became the focus of a substantial theoretical effort in the field, where we point to the whitepapers~\cite{Bauer2022, DiMeglio2023} that summarize the progress and lay out the roadmap. In this context, lattice gauge theory~\cite{Wilson1974} (LGT) furnishes a non-perturbative approach to many problems at hand. Even though the Hamiltonian formulation was derived very early~\cite{Kogut1975}, its application to universal quantum computation was realized and examined under this aspect much later~\cite{Byrnes2005}. By simple considerations it became clear that contrary to the use of LGT on classical computers, which is often hampered by the fermionic sign problem, the mapping of the bosonic degrees of freedom (dof) to the quantum computational dof requires truncation. However, the approximation involved is \emph{controllable} and systematic error bounds e.g. for Hamiltonian time evolution can be derived~\cite{Tong2021}. Given this solid theoretical ground, the development of efficient formulations for non-Abelian LGT is hence of great interest~\cite{Davoudi2020} and ranges from discrete subgroups~\cite{Assi2024} for pure gauge theory to formulations in terms of gauge invariant dof s.a. the loop-string-hadron formulation~\cite{Davoudi2023} for theories with matter.

In this work we investigate another aspect of the mapping of lattice gauge theory and estimate the resource requirements if we vary the standard computational unit. In particular, besides two-level qubits we consider multi-level qudits with a state space of $d>2$~\cite{Wang2020} and quantum modes (qumodes)~\cite{Weedbrook2012} as used for continuous-variable quantum computation~\cite{Lloyd1999} on photonic devices. 

Testing ground for our study will be the strong-coupling limit of lattice QCD as re-examined recently in~\cite{scnf12023}. Actual interest in this theory arose with the work of~\cite{Rossi1984, Karsch1989}, back in the early days of Euclidean LGT, mainly motivated by the fact that the fermionic sign problem of the theory remained tractable to some degree. As a consequence the phase diagram of the theory with $N_f=1$ staggered quark flavor could be obtained~\cite{Forcrand2010}, soon followed by the same study including gauge corrections~\cite{Forcrand2014}. 

By taking the continuous-time limit of the Euclidean LGT, the Hamiltonian for the theory with varying number of quark flavors could be constructed~\cite{Klegrewe2020, WUproc2021, WUproc2022}. It is this last development which makes the theory interesting from the perspective of quantum simulation: The resulting Hilbert space is a discrete state-space of physical, color-singlet dof (mesons and baryons) whose interaction is controlled by a local, sparse nearest-neighbor Hamiltonian. However, the Hilbert space dimension of the theory grows quickly with increasing number of staggered quark flavors $N_f$ and is thus quantum computationally well beyond the hardware limits of the NISQ era. For this reason we study the theory and its resource requirements after a suitable map onto different computational units (qubit, qudit or qumode). Before we lay out the results of this analysis in Sect.\ref{sec:resu}, we present a short account of the derivation of the strong coupling theory in the Hamiltonian formulation in the following Sect.\ref{sec:background}.

%% file: Background.tex
\section{Background}
\label{sec:background}
\noindent In Ref.\cite{Klegrewe2020} the dual formulation of strong coupling Lattice QCD with $N_f=1$ flavors of staggered quarks was derived in the limit of continuous Euclidean time by letting the lattice extent in time direction $N_\tau\rightarrow\infty$, while keeping the temperature in spatial lattice units $aT \sim \gamma^2/N_t$ fixed, where $\gamma$ represents the bare anisotropy parameter. The resulting partition function at temperature $T$ and baryon chemical potential $\mu_B$,
\begin{align}
\calZ_{\rm CT}(aT,a\mu_B)&=\Tr_\Hil\left[e^{(-\hat{\calH}+\hat{\calN}a\mu_B)/aT}\right],
\label{eq:CTParFunc}
\end{align}
gives then rise to a Hamiltonian and number operator
\begin{align}
\hat{\calH}&=-\frac{1}{2}\sum_{
\langle x,y\rangle}
\left(
\hat{J}^{+}_{x}\hat{J}^{-}_{y}+
\hat{J}^{-}_{x} \hat{J}^{+}_{y}
\right),&\hat{\calN}&=\sum_{x}\hat{\omega}_x.
\label{eq:hamiltonian}
\end{align}
with matrix representations
\begin{align}
\hat{J}^+_x&=\left(
\begin{array}{cccc|cc}
0   & 0   & 0   & 0 &  & \\
v_L & 0   & 0   & 0 &  & \\
0   & v_T & 0   & 0 &  & \\
0   & 0   & v_L & 0 &  & \\
\hline
 &  &  &  & 0 & 0\\
 &  &  &  & 0 & 0\\
\end{array}
\right),\,
\hat{\omega}=\left(
\begin{array}{cccc|cc}
0 & 0 & 0 & 0 &  & \\
0 & 0 & 0 & 0 &  & \\
0 & 0 & 0 & 0 &  & \\
0 & 0 & 0 & 0 &  & \\
\hline
 &  &  &  & 1 & 0\\
 &  &  &  & 0 & -1
 \label{eq:Js}
\end{array}
\right).
\end{align}
The trace in~\eref{eq:CTParFunc} is over the Hilbert space $\Hil=\bigotimes_{x\in\Sigma} \Hil_{x}$ where $\Hil_{x}$ is six dimensional, with states $|\hadron\rangle=| \meson,\baryon\rangle\in \Hil_{x}$ and a basis $|\hadron_i\rangle$ of six hadronic states (four mesonic and two baronic) per spatial lattice site, 
\begin{align}
|\hadron_i\rangle \in \{0,\pi,2\pi,3\pi, B^+, B^-\}\,.
\label{eq:hadron_state}
\end{align}
With the mesonic occupation number $\meson = 0, \pi, 2\pi, 3\pi$ being also bounded from above, the theory has a particle-hole symmetry. If we let $\spin = \meson-3/2$, we see that the mesonic part of the Hamiltonian in Eq.(\ref{eq:hamiltonian}) actually corresponds to the quantum Heisenberg model in spin $j = N_c/2 = 3/2$ representation with states $|j,\spin\rangle$.\footnote{We note that the antiferromagnetic Heisenberg Hamiltonian also emerges as an effective theory of Lattice QCD in the strong coupling limit in lowest order of Hamiltonian perturbation theory~\cite{Smit1980}.}

\noindent Also, owing to the limit of infinite bare gauge coupling, $\beta=2N_c/g^2 = 0$, baryons become static with trivial dynamics in the continuous time limit - a fact, that allows for importance sampling of the theory at finite baryon density \emph{without} a sign problem. As a consequence, $\hat{\calH}$ and $\hat{\calN}$ in Eq.(\ref{eq:hamiltonian}) have a block-diagonal structure and the states $\ket{\hadron}_x$ of the local Hilbert space are a direct \textit{sum} of mesonic and baryonic occupational states, $\ket{\hadron}_x = \ket{\meson}_x \oplus \ket{\baryon}_x$.

\begin{figure*}[ht]
\centering
\subfigure[]{
\centering
\includegraphics[scale=1.6]{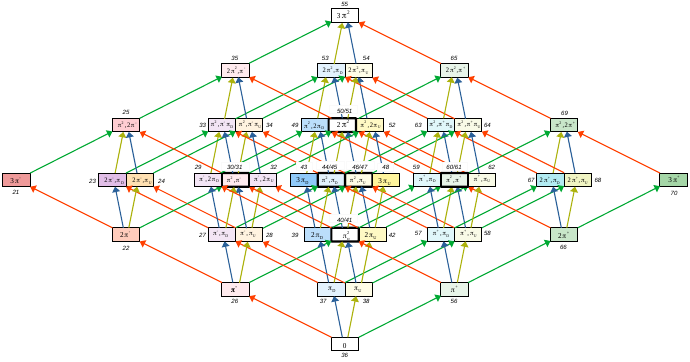}
\label{fig:transitionsB0}
}

\subfigure[]{
\includegraphics[scale=1.0]{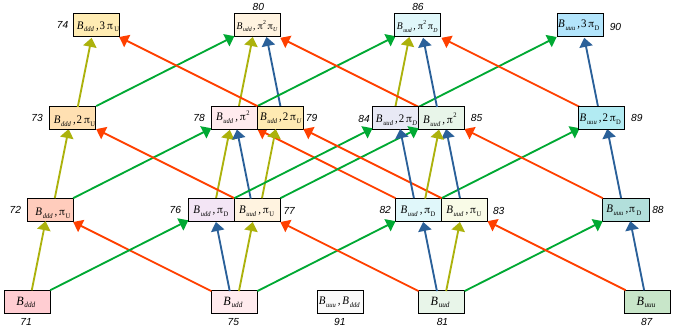}
\label{fig:transitionsB1}
}

\subfigure[]{
\includegraphics[scale=1.0]{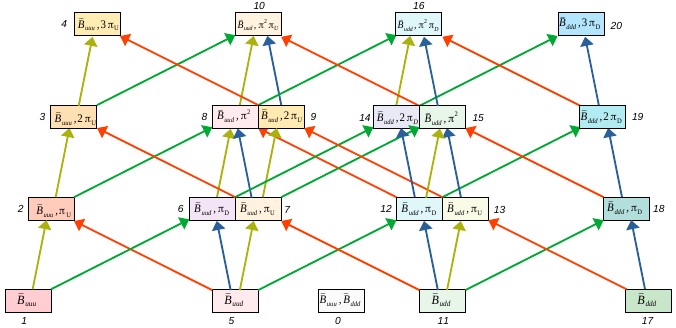}
\label{fig:transitionsBm1}
}

\caption{\label{fig:nf2_states}Hadronic states in the sectors with baryon number $B = 0$ $(a)$, $B = 1,2$ $(b)$ and $B=-1,-2$ $(c)$. The action of the ladder operators $J^+_{Q_i}$ on a state is indicated by yellow, blue, green and red arrows for the conserved currents $Q_i = \pi_u,\pi_d, \pi^+$ and $\pi^-$, respectively. In each sector, Isospin $I$ increases horizontally from left to right, the mesonic occupation number $\spin$ increases vertically, bottom to top.}
\end{figure*}

In~\cite{WUproc2021, WUproc2022, WUproc2023} the theory was extended to $N_f=2$ flavors of staggered quarks. The partition function is now given by
\begin{align}
Z_{\rm CT}(\bareT,\bareMu,\bareI)&=
\Tr_\Hil\left[e^{(-\hat{\calH}+\hat{\calN}_B\bareMu+\hat{\calN}_I\bareI)/\bareT}\right]
\label{eq:2flavor_partition_function}
\end{align}
where the Hamiltonian is a sum of four contributions, one for each ``pion''
$\pi\in\{\pi^+,\pi^-,\pi_u,\pi_d\}$:
\begin{align}
  \hat{\calH}&=-\frac{1}{2}\sum_{\langle x,y\rangle}
 \sum_{Q_i \in \{ \substack{\pi^+, \pi^-,\\ \pi_U, \pi_D} \}} \lr{
 {\hat{J}_{Q_i,x}}^+ {\hat{J}_{Q_i,y}}^- + {\hat{J}_{Q_i,x}}^- {\hat{J}_{Q_i,y}}^+}.
 \label{eq:hamiltonian_nf2}
 \end{align}
In addition to the baryon number operator $\hat{\calN}_B = {\rm diag}(-2,-1,\ldots 1,2)$, the isospin number operator $\hat{\calN}_I={\rm diag}\lr{0,-\frac{3}{2},\ldots \frac{3}{2},0}$ has been introduced and coupled to the isospin chemical potential $a\mu_I$. The local Hilbert space $\Hil_{x}$ now has a basis of 92 distinct hadronic states $|\hadron\rangle$. The particle-hole symmetry of the one-flavor case with quantum number $\spin$ generalizes to $\spin = \meson - \frac{N_c}{2}(N_f-|B|)$, in addition the states can be labelled by the local baryon number $B$ and isospin $I$ as shown in Table \ref{HadronStatesNf2}. Once again, as the baryon number is conserved and baryons remain static, the ladder operators $\hat{J}_{Q_i,x}^{+/-}$ in Eq.(\ref{eq:hamiltonian_nf2}) have block-diagonal structure and decompose into sectors of fixed local baryon number $B=-2,\ldots, 2$. In each sector the ladder operators generate a corresponding state multiplet which is visible in Fig.\ref{fig:transitionsB0}-Fig.\ref{fig:transitionsBm1}. The action of the ladder operators $J^+_{Q_i}$ on a state $|\hadron\rangle$ is indicated by yellow, blue, green and red arrows for the conserved currents $Q_i = \pi_u,\pi_d, \pi^+$ and $\pi^-$, respectively. Apart from the state number, a state is labelled by its mesonic and baryonic occupation number where we point to~\cite{WUproc2021, WUproc2022} for a full account of the derivation of the matrix elements and the state definition. For our purpose, it suffices to know that the $\hat J^+_{Q_i}$ once again generate a representation of a $SU(2)$ algebra, in particular
\beq
\frac{N_c}{2}[\hat J^+_{Q_i},\hat J^-_{Q_j}] = \hat J^3_{Q_i} \delta_{ij}\,.
\eeq
This time however the nature of the representation is more intricate. With the particular labelling chosen in Fig.\ref{fig:transitionsB0}-Fig.\ref{fig:transitionsBm1}, where isospin $I$ increases horizontally from left to right and $\spin$ increases vertically, c.f. Table \ref{HadronStatesNf2}, we see that the representation of the $J^+_{\pi_u}$ and $J^+_{\pi_d}$, which are isospin conserving, is again reducible and hence becomes block-diagonal in each isospin sector, where it in turn forms a product representation.

\begin{table*}
\begin{center}
\begin{tabular}{|r|r||c|c|c|c|c|c|c|c|c|c|c|c|c||c|}
\hline
$B$ & $I$\; & \multicolumn{13}{c||}{$\spin=\meson-\frac{3}{2}(2-|B|)$} & $\Sigma$ \\
\hline
 &  & $-3$ & $-\frac{5}{2}$ & $-2$ & $-\frac{3}{2}$ & $-1$ & $-\frac{1}{2}$ & $\, 0\,$ & $+\frac{1}{2}$ & $+1$ & $+\frac{3}{2}$ & $+2$ & $+\frac{5}{2}$ & $+3$ &  \\
\hline
\hline
 -2& 0 & & & & & & & 1 & & & & & & & 1 \\
\hline
 -1 & $-\frac{3}{2}$ &&&& 1 &&  1 && 1 && 1 &&&&  4 \\
 -1 & $-\frac{1}{2}$ &&&& 1 &&  2 && 2 && 1 &&&&  6 \\
 -1 & $+\frac{1}{2}$ &&&& 1 &&  2 && 2 && 1 &&&&  6 \\
 -1 & $+\frac{3}{2}$ &&&& 1 &&  1 && 1 && 1 &&&&  4 \\
\hline 
0 & $-3$ &  &&   &&& & 1 &&&&&&& 1\\ 
0 & $-2$ &  &&   && 1 && 2 && 1 && & && 4\\
0 & $-1$ &  && 1 && 2 && 4 && 2 &&  1 && & 10 \\
0 & $0$  & 1&& 2 && 4 && 6 && 4 && 2 && 1 & 20 \\
0 & $+1$ &  && 1 && 2 && 4 && 2 &&  1 && & 10 \\
0 & $+2$ &  &&   && 1 && 2 && 1 && & && 4\\
0 & $+3$ &  &&   &&& & 1 &&&&&&& 1\\ 
\hline
 1 & $-\frac{3}{2}$ &&&& 1 &&  1 && 1 && 1 &&&&  4 \\
 1 & $-\frac{1}{2}$ &&&& 1 &&  2 && 2 && 1 &&&&  6 \\
 1 & $+\frac{1}{2}$ &&&& 1 &&  2 && 2 && 1 &&&&  6 \\
 1 & $+\frac{3}{2}$ &&&& 1 &&  1 && 1 && 1 &&&&  4 \\
\hline
 2 & 0 &&&&&&& 1 & & & & & & & 1 \\
\hline
\hline
$\Sigma$ &  &1 & 0 & 4 & 8 & 10 & 12 & 22 & 12 & 10 & 8 & 4 & 0 & 1 & 92\\    
\hline
\end{tabular}
\end{center}
\caption{
State multiplicities of the local state space for the two-flavor theory in the Hamiltonian formulation, totalling $92$ states at site $x$. The states are labelled by baryon number $B$, isospin number $I$ and symmetrized meson occupation number $\spin=\meson-\frac{\Nc}{2}(\Nf-|B|)$.}
\label{HadronStatesNf2}
\end{table*}

%% file: Results.tex
\section{Results}
\label{sec:resu}
\subsection{$N_f = 1$}
\subsubsection{General Remarks}
\label{subsubsec:nf_1_qubits}
 \begin{figure}[ht]
\centering
\includegraphics[scale=.3]{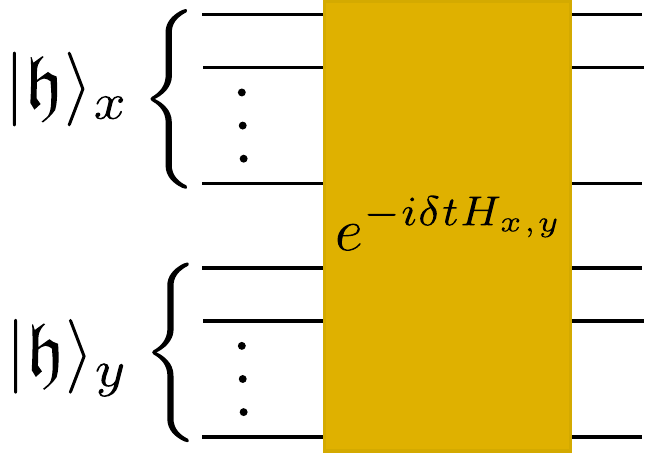}
\caption{\label{fig:block_Trotter} Schematic of a single Trotter step of step size $\delta t$, connecting the $nn$-sites $x$ and $y$ with local states $|\hadron\rangle_x$ and $|\hadron\rangle_y$.}
\end{figure}
The operators appearing in Eq.(\ref{eq:Js}) have block-diagonal structure: The meson and baryon sector do not mix when acting with the Hamiltonian on the product space $\Hil_{x}\otimes \Hil_{y}$. If we extend the $|B|>0$ sector also to 4 dimensions, we can write the one-flavor Hamiltonian in terms of $|B|$-projectors,
\begin{align}
\hat{\calH}=-\frac{1}{2}\sum_{
\langle\vec{x},\vec{y}\rangle}
|0\rangle\langle 0|_x |0\rangle\langle 0|_y\left(
\hat{J}^{+}_{\vec{x}}\hat{J}^{-}_{\vec{y}}+
\hat{J}^{-}_{\vec{x}} \hat{J}^{+}_{\vec{y}}
\right)\nonumber\\ + a\mu_B\sum_{\vec{x}} |1\rangle\langle 1|_x\hat{\omega}_x.
\label{eq:hamiltonian_proj}
\end{align}
We used that property to derive the gate set for $N_f=1$ in terms of controlled gates where locally this amounts to an ancillary bit, controlling the baryonic sector~\cite{scnf12023}.

\subsubsection{Mapping to Qudits}
Using a $d$-level systems with $d>2$ (qudit) as quantum computational unit, the dimension of the local Hilbert space in Eq.(\ref{eq:hadron_state}) suggests the use of qudits with $d=6$. Rewriting the Hamiltonian in terms of $\hat J^1 = \frac{\sqrt{3}}{2}(J^+ + J^-)$ and $\hat J^2 = \frac{\sqrt{3}}{2i}(\hat J^+ - \hat J^-)$, the $\hat J^i$ can be decomposed into generators $\mathcal{X}_{ij}, \mathcal{Y}_{ij}$ of unitary Givens rotations $G_{jk}(\theta, \phi)$, with
\beq
G_{jk}(\theta, \phi) = e^{-i\theta(\cos{\phi}\mathcal{X}_{jk}-\sin{\phi}\mathcal{Y}_{jk})}\, ,
\eeq
where~\cite{OLeary2006}
\beq
\mathcal{X}_{kl} = |k\rangle\langle l | + | l\rangle\langle k |,\,\mathcal{Y}_{kl} = -i (|k\rangle\langle l | - | l\rangle\langle k |)\, .
\label{eq:generalized_XY}
\eeq
One obtains
\beq
 J^1 &=& \frac{\sqrt{3}}{2}(\mathcal{X}_{01}+\frac{2}{\sqrt{3}}\mathcal{X}_{12}+\mathcal{X}_{23})\nonumber\\ 
 J^2 &=& -\frac{\sqrt{3}}{2}(\mathcal{Y}_{01}+\frac{2}{\sqrt{3}}\mathcal{Y}_{12}+\mathcal{Y}_{23})\, .
 \label{eq:givens_decompo}
 \eeq
If we trotterize the time evolution for a pair of neighboring mesonic sites $x,y$ with the hopping term in Eq.(\ref{eq:hamiltonian}) according to
\beq
e^{-i\delta t H_{x,y}} = \prod_i U_{\mathcal{XX}_i}U_{\mathcal{YY}_i} + \mathcal{O}(\delta t^2)\,
\eeq
we obtain a decomposition into 9 two-qudit entangling operations of the generic form $U_{\mathcal{XX}} = e^{-i \alpha \mathcal{X}^x_{kl}\otimes \mathcal{X}^y_{mn}}$ and $U_{\mathcal{YY}} = e^{-i \beta \mathcal{Y}^x_{kl}\otimes \mathcal{Y}^y_{mn}}$, respectively. These elementary operations can be realized via phase-compensated M\o lmer S\o rensen (MS) gates for trapped-ion qudits~\cite{Molmer1998, Ringbauer2021}. In~\cite{Low2019} a generalization of the MS gate was proposed that allows for the simultaneous driving of multiple transitions between qudit levels of the involved neighboring qudits in an entangling operation. In other words, a linear combination of two-level transitions s.a.~Eqs.(\ref{eq:givens_decompo}) can be realized by just one operation.  As a result, the hopping term can be implemented via two two-qudit entangling-operations
\beq
e^{-i\delta t H_{x,y}} = e^{-i \delta t J^1_{x}\otimes J^1_{y}} e^{-i \delta t J^2_{x}\otimes J^2_{y} }+ \mathcal{O}(\delta t^2),\,
\label{eq:time_evo_nf1_qudit}
\eeq
thus reducing the gate depth to 2 for the $nn$-term. This substantial reduction in entangling operations comes at the cost of increased complexity in experimental control, where we refer to the analysis carried out in~\cite{Calajo2024} to address the experimental challenges.\footnote{See also~\cite{Illa2024} for a recent resource analysis of digital quantum simulations of (1+1)D Lattice QCD on $d=8$ trapped ion qudits.}

\noindent In contrast, the number operator $\hat{\calN}$ in Eq.(\ref{eq:hamiltonian}) just corresponds to a single, diagonal qudit operation. Picking as a basis for diagonal gates the $\mathcal{Z}_{jk}$ with
\beq
\mathcal{Z}_{jl} = |j\rangle\langle j | - | l\rangle\langle l |
\eeq
that, together with $\mathcal{X}_{jl}$ and $\mathcal{Y}_{jl}$, generate a $su(2)$ subalgebra of $SU(d=6)$, the local term in Eq.(\ref{eq:hamiltonian}) then generates an one-qudit operation of the type $e^{-i t \mu_B\mathcal{Z}_{45}}$ for each qudit during time evolution. As this corresponds to a single diagonal gate without Trotterization, the main cost of the time evolution will be given by Eq.(\ref{eq:time_evo_nf1_qudit}). 
\subsubsection{Mapping to Qumodes}
\label{subsec:nf1_qumodes}
\noindent As a third alternative, we can encode our system in the degrees of freedom of quantum harmonic oscillators, i.e. qumodes. The basic operator set then consists of the quadrature variables $(\hat{x_i},\hat{p_i})$ or, alternatively, Fock-space creation and annihilation operators, $\hat a_i, \hat a_i^\dagger$, with the usual relations
\beq
\hat x_i &=& \frac{1}{2}(\hat a_i+\hat a_i^\dagger),\,\hat p_i = \frac{1}{2i}(\hat a_i-\hat a_i^\dagger)\label{eq:quadratures}\\
\hat a_i &=& x_i+ip_i,\, a_i^\dagger = x_i-ip_i
\eeq
and
\beq
[\hat a_i,\hat a_j] = [\hat a_i^\dagger,\hat a_j^\dagger] = 0, [\hat a_i,\hat a^\dagger_j] = \delta_{ij}, {[\hat x_i, \hat p_j]} = \frac{i}{2}\delta_{ij}\,.\nonumber
\eeq
The $j=\frac{N_c}{2}$ representation of the $J^+$ and $J^-$ (non-trivial block in Eq.(\ref{eq:Js})) can be obtained via the Jordan-Schwinger-map~\cite{Schwinger1952, Annabestani2020}
\beq
\hat J_x^+ = \hat a_{1,x}^\dagger \hat a_{2,x},\,\hat J_x^- = \hat a_{2,x}^\dagger \hat a_{1,x},
\eeq
s.t.
\beq
|j,m\rangle = \frac{(\hat  a_1^\dagger)^{j+m}(\hat  a_2^\dagger)^{j-m}}{\sqrt{(j+m)!(j-m)!}}|0\rangle\,.
\eeq
The local number operators $\hat n_i = \hat a_i^\dagger \hat a_i,\, i=1,2$ then have eigenvalues $n_{1/2}$ which are constrained to $n_1+n_2 = 2j = N_c$. The mesonic hopping term of the Hamiltonian Eq.(\ref{eq:hamiltonian}) hence reads
\beq
\calH_{meson} = -\frac{1}{2}\sum_{
\langle{x},{y}\rangle}(\hat  a^\dagger_{1,x}\hat a_{2,x}\hat a^\dagger_{2,y}\hat a_{1,y}) + h.c.\,.
\label{eq:h_meson_in_bosonic_ops}
\eeq
Expressed in the quadrature field operators Eq.(\ref{eq:quadratures}) this expression becomes
\beq
\calH_{meson} = -\frac{1}{2}\sum_{
\langle{x},{y}\rangle}\sum_{\{i_k\}}(\hat  q^{i_1}_{1,x}\hat q^{i_2}_{2,x}\hat q^{i_3}_{2,y}\hat q^{i_4}_{1,y}) s_{\{i_1,\ldots, i_4\}},
\label{eq:h_meson_in_quadratures}
\eeq
with the shorthand $(\hat q^1,\hat q^2) = (\hat x, \hat p)$ and signs $s_{\{i_1,\ldots, i_4\}}$\footnote{$s_{\{2,1,1,2\}} = s_{\{1,2,2,1\}} = -1$, otherwise $s_{\{i_1,\ldots, i_4\}} = 1$.}. In the following we assume to have access to the qumode instruction set
\beq
\{e^{is\hat x_k}, e^{is\hat a_k^\dagger \hat a_k}, e^{is\hat x_k^2}, e^{is\hat x_k^3}, e^{is\hat x_k \hat x_j}\}\nonumber\\
\equiv\{Z(s), R(s), P(s), V(s), CZ(s)\}
\label{eq:inst_set}
\eeq
In particular, the Fourier gate $F \equiv e^{i\frac{\pi}{2}\hat a_k^\dagger \hat a_k}$ can be used for unitary conjugation of powers of quadrature operators, e.g. $g(\hat p_k) = F_k g(\hat x_k) F_k^\dagger$ for polynomials $g$. This allows us to rewrite each of the eight terms in Eq.(\ref{eq:h_meson_in_quadratures}) in the form $\hat x_{1,x} \hat x_{2,x} \hat x_{1,y} \hat x_{2,y}$. Through exact gate decompositions (Appendix \ref{subsec:exact_decompo}) trotterized time evolution by a single term $e^{i\delta t \hat x_{1,x} \hat x_{2,x} \hat x_{1,y} \hat x_{2,y}}$ is found to be costly, involving approximately $8\times 65$ elementary operations of the instruction set Eq.(\ref{eq:inst_set}), which reduces to $8\times 11$ elementary operations, depending on the availability of a quartic gate $Q(s) = e^{is\hat x_k^4}$. In total, a first-order Trotter step of a single mesonic $nn$-pair, Eq.(\ref{eq:meson_trotter_qumode}), results in a gate count of $64 Q, 147CZ, 149F$ and a corresponding gate depth of 348 (see Appendix \ref{subsec:exact_decompo}), now under the assumption that the quartic gate is part of the gate set.

Similarily, the baryonic term of Eq.(\ref{eq:hamiltonian}) can be mapped to 
\beq
\calH_{baryon} = \sum_x(\hat n_{1,x}-\hat n_{2,x}) = \sum_{x,i}(\hat x^2_{i,x}+\hat p_{i,x}^2)
\label{eq:h_baryon_in_bosonic_ops}
\eeq
now applied to Fock-states with the constraint $n_1+n_2 = 1$. The exponentiated term can be implemented using two rotation gates $R(\mu_B\delta t) = e^{i\mu_B \delta t \hat a_k^\dagger \hat a_k}$ per baryonic site $x$.

In order to combine the mesonic and baryonic time evolution, we can follow the strategy of Sect.\ref{subsubsec:nf_1_qubits} and control the local sector (mesonic or baryonic) via an ancillary qubit, yielding a Hamiltonian
\beq
\calH = -&\frac{1}{2}&\sum_{
\langle{x},{y}\rangle}|0\rangle \langle 0|_x|0\rangle \langle 0|_y\sum_{\{i_k\}}(\hat  q^{i_1}_{1,x}\hat q^{i_2}_{2,x}\hat q^{i_3}_{2,y}\hat q^{i_4}_{1,y}) s_{\{i_1\ldots i_4\}}\nonumber\\
&+& a\mu_B\sum_{x} |1\rangle\langle 1|_x(\hat n_{1,x}-\hat n_{2,x})
\label{eq:h_in_qumodes}
\eeq
Such control implies a mixed qubit-qumode hardware architecture as used in circuit QED (cQED)~\cite{Blais2020}. 

Alternatively, we could use the fact that either part of the Hamiltonian, $\calH_{meson}$ and $\calH_{baryon}$, commutes with the local number operator $\hat n = \hat n_1+ \hat n_2$ and hence does not change the local occupation number $n = n_1+ n_2 =  2j$ of a single Fock-state. Consequently, we can use the number operator to control the application of the time evolution operator in the baryonic and mesonic sector \emph{without} ancillary qubits, by letting e.g.
\beq
&\calH_{meson}& = -\frac{1}{2}\sum_{
\langle{x},{y}\rangle} H_{x,y}\frac{1}{2}(\hat n_x -1)\otimes \frac{1}{2}(\hat n_y -1)\nonumber\\
&\calH_{baryon}& = a\mu_B\sum_{x} \mathcal{N}_x \frac{1}{2}(3-\hat n_x)\,,
\label{eq:n_controlled_hams}
\eeq
in a manner reminiscent of the definition of controlled qumode operations s.a. the controlled Kerr gate~\cite{Killoran2019}. In the above Eq.(\ref{eq:n_controlled_hams}) we have implicitly defined the local hopping term $H_{xy}$ and on-site terms $\mathcal{N}_x$ from Eqs.(\ref{eq:h_meson_in_bosonic_ops}) and (\ref{eq:h_baryon_in_bosonic_ops}), respectively. We note that the terms in Eq.(\ref{eq:n_controlled_hams}) then actually correspond to a higher order interaction. In view of the already considerable cost to trotterize a time evolution with Eq.(\ref{eq:h_meson_in_bosonic_ops}), we have not pursued this further.

\subsection{$N_f = 2$}
\subsubsection{General remarks, mapping to qubits}
\noindent When mapping the two-flavor theory to qubits, we can use the block-diagonal shape of the ladder operators, c.f. Fig.\ref{fig:nf2_states}, and rewrite the Hamiltonian Eq.(\ref{eq:hamiltonian_nf2}) in terms of projectors of baryonic sectors, in complete analogy to the one-flavor case, Eq.(\ref{eq:hamiltonian_proj}):
\begin{align}
  \hat{\calH}&=-\frac{1}{2}\sum_{\langle x,y\rangle}\sum_{i,j}
  |i\rangle\langle i|_x |j\rangle\langle j|_y
 \sum_{Q_k} \lr{
 {\hat{J}_{Q_k,x}}^{(i)+} {\hat{J}_{Q_k,y}}^{(j)-} + {\hat{J}_{Q_k,x}}^{(i),-} {\hat{J}_{Q_k,y}}^{(j),+}%
 }
 \label{eq:hamiltonian_nf2_proj}
 \end{align}
In the above expression we have expanded each sector's dimension by "zero padding" to match the dimensionality of the largest sector ($B=0$) which is 50-dimensional. Formally, the expression Eq.(\ref{eq:hamiltonian_nf2_proj}) can be derived by noting that locally for $x \in\Sigma$ we have
 $J^+_{Q_k,x} = \sum_i |i\rangle\langle i|_x \otimes J^{(i)+}_{Q_k,x}$ where we wrote the Kronecker product explicitly. In the product $ {\hat{J}_{Q_k,x}}^+ \otimes{\hat{J}_{Q_k,y}}^-$ we then use the permutation equivalence $A\otimes B = P B\otimes A P^T$ ($A,B$ square matrices) to order the projectors $ |i\rangle\langle i|_x |j\rangle\langle j|_y$ to the left,
 \begin{align}
  {\hat{J}_{Q_k,\vec{x}}}^+ &\otimes\hat{J}_{Q_k,\vec{y}}^- = \sum_{i,j} |i\rangle\langle i|_x \otimes J^{(i)+}_{Q_k,x}\otimes |j\rangle\langle j|_y \otimes J^{(j)-}_{Q_k,y}\nonumber\\
  &= \sum_{i,j} |i\rangle\langle i|_x \otimes P^T P J^{(i)+}_{Q_k,x}\otimes |j\rangle\langle j|_y P^T P\otimes J^{(j)-}_{Q_k,y}\nonumber\\
  &=  P^T\sum_{i,j} |i\rangle\langle i|_x \otimes |j\rangle\langle j|_y \otimes J^{(i)+}_{Q_k,x} \otimes J^{(j)-}_{Q_k,y} P\,.
 \end{align}
 The permutation matrix $P$ then acts on the product state $\in \Hil_{x}\otimes \Hil_{y}$ as $P(|i\rangle_x|\hadron_i\rangle_x |j\rangle_y|\hadron_j\rangle_y) = |i\rangle_x |j\rangle_y
 |\hadron_i\rangle_x |\hadron_j\rangle_y$. It is in this sense that Hamiltonian Eq.(\ref{eq:hamiltonian_nf2_proj}) is equivalent to Eq.(\ref{eq:hamiltonian_nf2}).\footnote{We note that the block-diagonal shape of the local Hamiltonian Eq.(\ref{eq:hamiltonian_nf2_proj}) may be interesting to study on its own: In Ref.\cite{Gu2021} such Hamiltonians are considered for fast-forwarding time evolution, i.e. a time evolution whose gate complexity is sublinear in $t$.}
 
\begin{figure}
\centering
\includegraphics[scale=.5]{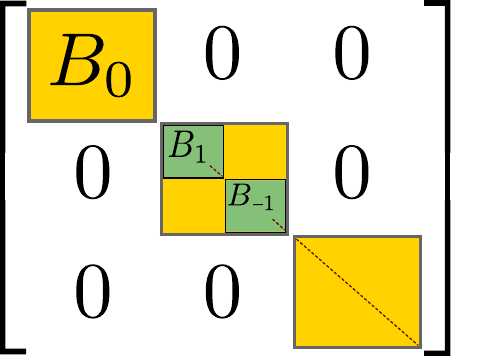}
\caption{\label{fig:block_J} Block-diagonal structure of the ladder operators $J_{Q_k}^{+/-}$ after reordering the basis and zero-padding (red dashed line).}
\end{figure}

\noindent By reordering the basis, the ancillary register $|i\rangle$ in Eq.(\ref{eq:hamiltonian_nf2_proj}) can be chosen to just encode the sectors $|B| = 0,1,2$. In Fig.\ref{fig:block_J} we show schematically the block-diagonal structure of the ladder operators, after reordering and zero-padding.
As practically we will use the ancillary states as control state, a unary encoding of the 3 states seems advisable, thus demanding 3 ancillary bits per site $x$. The actual state at this site will then be encoded using 6 qubits, hence totalling 9 qubits to represent $|\hadron\rangle_x$. For the sector $i = |B| = 1$, we can decompose the $J_{Q_k}^{(i)+/-}$ even further (c.f. Fig.\ref{fig:block_J}) by noticing that
 \begin{align}
 J_{Q_k}^{(|B|=1)-/+} = |0\rangle\langle 0| \otimes J_{Q_k}^{(B=-1)-/+} + |1\rangle\langle 1| \otimes J_{Q_k}^{(B=1)-/+}\,.
 \end{align}
Because of the $J_{Q_k}^{(B=-1)-/+} = J_{Q_k}^{(B=1)-/+}$, the qubit count of 6 ``substate'' qubits will be reduced by one, ultimately resulting in a lower gate count.
 \begin{figure}
\centering
\includegraphics[scale=.5]{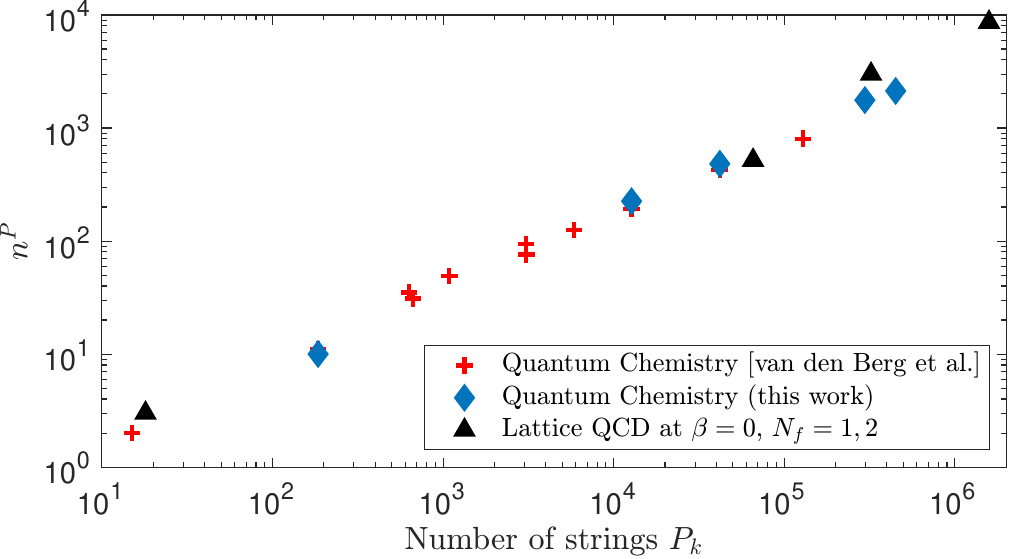}
\caption{\label{fig:fam_vs_string_num} The number of partitions $n^P$ for different number Pauli strings $P_k$ in our theory, using a sequential approach. Also shown are the results obtained for a selected set of quantum chemistry Hamiltonians, along with data taken from~\cite{qiskitnature2023} (Table 3) on the same set, for comparison.}
\end{figure}

 \begin{figure}
\centering
\includegraphics[scale=.5]{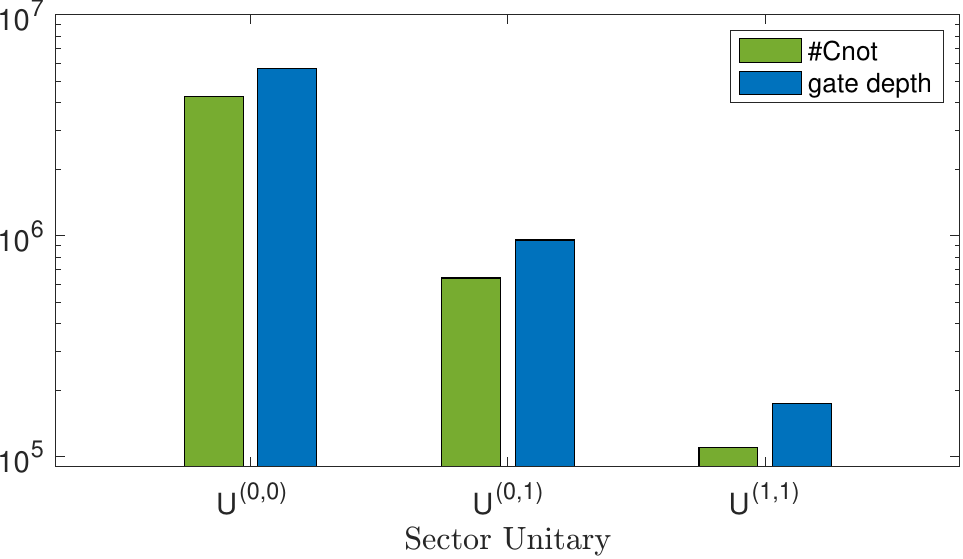}
\caption{\label{fig:nf2_qubit_gc} The number of entangling CNOT gates and corresponding gate depth for the circuits implementing the unitaries Eq.(\ref{eq:nf2_decompo_qubit}) in the baryonic sectors $(i,j)$ at sites $x$ and $y$, respectively.}
\end{figure}
\noindent Finally, to obtain the circuit depth for the time evolution with the Hamiltonian Eq.(\ref{eq:hamiltonian_nf2_proj}), we have to diagonalize and optimize each of the factors 
 \begin{align}
e^{-i\delta t H_{x,y}} &= e^{i\frac{\delta t}{2} \sum_{ij}|i\rangle\langle i|_x |j\rangle\langle j|_y
 \sum_{Q_k} \lr{
 {\hat{J}_{Q_k,x}}^{(i)+} {\hat{J}_{Q_k,y}}^{(j)-} + {\hat{J}_{Q_k,x}}^{(i),-} {\hat{J}_{Q_k,y}}^{(j),+}}} \nonumber\\
 &=\prod_{ij} |i\rangle\langle i|_x |j\rangle\langle j|_y \nonumber e^{i\frac{\delta t}{2} \sum_{Q_k} \lr{
 {\hat{J}_{Q_k,x}}^{(i)+} {\hat{J}_{Q_k,y}}^{(j)-} + {\hat{J}_{Q_k,x}}^{(i),-} {\hat{J}_{Q_k,y}}^{(j),+}}}\\
 &\equiv U^{(0,0)}U^{(1,0)}U^{(0,1)}U^{(1,1)}\,
 \label{eq:nf2_decompo_qubit}
 \end{align}
 where the individual Trotter factors for different sectors commute (orthogonality). The $U^{(i,j)}$-diagonalization can be done using the methods presented in~\cite{Berg2020, Murairi2022}. Given the problem size involving a high dimensional Hamiltonian with 6-qubit-to-6-qubit coupling as can be seen in Eq.(\ref{eq:nf2_decompo_qubit}), a challenge that arises during the optimization is the partitioning into commuting sets of the large number of Pauli strings $P_k$ obtained from the expansion
 \beq
\sum_{Q_k} \lr{
 {\hat{J}_{Q_k,x}}^{(i)+} {\hat{J}_{Q_k,y}}^{(j)-} + {\hat{J}_{Q_k,x}}^{(i),-} {\hat{J}_{Q_k,y}}^{(j),+}} = \sum_l^{N_{ij}} c_l P_l\,.
 \eeq
With the number of strings $N_{ij}$ being of $\mathcal{O}(10^6)$ we chose a sequential approach~\cite{Berg2020} over graph coloring. Fig.\ref{fig:fam_vs_string_num} shows the resulting number of partitions $n^P_{ij}$ for the different $N_{ij}$ appearing in our problem, along with a selected set of partitions obtained from quantum chemistry Hamiltonians available from~\cite{qiskitnature2023}. Unsurprisingly, the $n^P_{ij}$ do not only depend on the number of Pauli strings (in the quantum chemistry problems shown the number of qubits varies between $4-48$), but also on their structure, which often involves identities $I$ for the molecules shown, thus increasing their ``commutativity''. \noindent Once the partitioning is obtained, we proceed with diagonalization and optimization~\cite{Berg2020, Murairi2022} and ultimately arrive at the circuits to simulate the sectors of the present theory, along with their gate count, using IBM Qiskit~\cite{Qiskit2023}. In Fig.\ref{fig:nf2_qubit_gc} we display the circuit depth and CNOT-count of the circuits implementing the unitary $U^{(i,j)}$ of Eq.(\ref{eq:nf2_decompo_qubit}), as expected the cost is largely dominated by the purely mesonic sector with unitary $U^{(0,0)}$ owing to its large dimension. This statement remains true, even if we include the parameter regime of $\mu_B,\mu_I>0$: The time evolution involving $\hat{\calN}_B$ and $\hat{\calN}_I$ (defined immediately below Eq.(\ref{eq:hamiltonian_nf2})) will again be controlled by the ancilla register encoding the baryonic sectors $|B|>0$. As both operators, $\hat{\calN}_B$ and $\hat{\calN}_I$, are diagonal in the chosen basis, they can be readily decomposed into sequences of CNOT gates and $R_Z$-gates of modest length, without Trotterization. The resource estimate given in Fig.\ref{fig:nf2_qubit_gc}, along with the qubit requirement of 9 qubits per site $x$ (3 ancillary qubits for an unary encoding of the baryonic sector and 6 qubits encoding the actual state) hence represent the central result of this section.
 
 \subsubsection{Mapping to qudits}
\noindent Turning to $d$-level systems with $d>2$, we can once again rewrite Eq.(\ref{eq:hamiltonian_nf2_proj}) in a form using the hermitian operators $J^{1/2}$ for each pion, to obtain
\begin{align}
  \hat{\calH}&\sim \sum_{\langle x,y\rangle}\sum_{i,j}
  |i\rangle\langle i|_x |j\rangle\langle j|_y
 \sum_{Q_k} \lr{
 {\hat{J}_{Q_{k,x}}}^{1,(i)} {\hat{J}_{Q_{k,y}}}^{1,(j)} + {\hat{J}_{Q_{k,x}}}^{2,(i)} {\hat{J}_{Q_{k,y}}}^{2,(j)}}.
 \label{eq:hamiltonian_nf2_proj_herm}
 \end{align}
The labelling of the baryon sectors $|B|=0,1,2$ then suggests a mixed qudit architecture as was investigated for state-preparation purposes in~\cite{Mato2024}, s.t.~one can use a qutrit as local ancilla controlling the baryonic sectors and a qudit with $d=50$ to encode the actual state at site $x$. To require a qudit of this high dimensionality may seem premature at first glance. However, recent progress in trapped-ion systems with \ce{^{137}Ba^{+}} as information carrier~\cite{Low2023} demonstrated control of a qudit with up to $d=13$ states, anticipating encodings up to $d=25$ which already corresponds to half the number of basis states in the $B=0$ sector of our theory and is capable to completely encode the $B>1$ sectors. Moreover, in~\cite{Dong2023} a photonic qudit with $d=25$ was realized using a cold-atom approach, showing that such high-dimensional Hilbert spaces can already be experimentally realized on different platforms. We hence anticipate the availability of qudit platforms with $d\lesssim 50$ for our purposes and favor this over the natural possibility of representing the $\mathcal{O}(50)$ states by the product states $|i_1\rangle \otimes |i_2\rangle$ of two qudits of lower dimensionality.

\noindent With the above specific architecture in mind, we trotterize the time evolution using the Hamiltonian Eq.(\ref{eq:hamiltonian_nf2_proj_herm}) by breaking up the exponential for each pion and represent the $J^{1,(i)}_{Q_k}$ and $J^{2,(i)}_{Q_k}$ in terms of the $\mathcal{X}$ and $\mathcal{Y}$, Eq.(\ref{eq:generalized_XY}), in complete analogy to the one-flavor case,
\beq
J^{1,(i)}_{Q_k} = \sum_l c_l \mathcal{X}_{m_ln_l},\, J^{2,(i)}_{Q_k} = \sum_l d_l \mathcal{Y}_{m_ln_l}, \forall\,i,k\,.
\label{eq:nf_j_decompo}
\eeq
Depending on the grouping of the $\mathcal{X}$ and $\mathcal{Y}$ into generalized MS gates, it is conceivable that the entangling gate count is reduced drastically in each sector combination $(i,j)$ when compared to the qubit case. After partial resummation of the $\mathcal{X}_{m_ln_l}$ and $\mathcal{Y}_{m_ln_l}$ in Eq.(\ref{eq:nf_j_decompo}) we obtain
\beq
J^{1,(i)}_{Q_k} = \sum^{N_i}_n \alpha^{(i,k)}_n A^{(i,k)}_n,\, J^{2,(i)}_{Q_k} = \sum^{N_i}_n \beta^{(i,k)}_n B^{(i,k)}_n\, ,
\eeq
where the $A_n$ and $B_n$ represent linear combinations (c.f. Eq.(\ref{eq:givens_decompo})) of the $\mathcal{X}$ and $\mathcal{Y}$, respectively. One mesonic Trotter step would hence involve the unitaries
\beq
&e&^{-i\delta t H_{x,y}} \approx \prod_{i,j} |i\rangle\langle i|_x |j\rangle\langle j|_y \times\nonumber\\ 
\prod_{k, n,m} &e&^{-i\delta t \alpha^{(i,k)}_n \alpha^{(j,k)}_m A^{(i,k)}_n\otimes A^{(j,k)}_m} e^{-i\delta t \beta^{(i,k)}_n \beta^{(j,k)}_m B^{(i,k)}_n\otimes B^{(j,k)}_m}\,.\nonumber
\eeq
Considering that, in the $|B|$=1 sector, there are 20 transitions per pion current $Q_k$, this suggests $N_1$ = 5 generalized MS gates à 4 transitions. Employing the same principle for $B = 0$, yields $N_0 = 7$ generalized MS gates to encode 28 transitions, c.f. Fig.\ref{fig:nf2_states}. In Table \ref{tab:nf2_qudit_gatecount} we summarize the arising gate count in terms of generalized MS-gates for fixed combination of baryonic sectors $(i,j)$ and fixed pion current $Q_k$.
\begin{table}[htp]
\begin{center}
\begin{tabular}{c|c|c}
Baryon Sector &0&1\\
\hline
0&$7\times 7\times 2$&$7\times 5\times 2$\\
1&$7\times 5\times 2$&$5\times 5\times 2$
\end{tabular}
\end{center}
\caption{ \label{tab:nf2_qudit_gatecount} Estimated count of generalized MS gates for the baryon sector combination $(i, j)$ per pion current $Q_k$. We assumed that the $28$ transitions in the $B = 0$ sector can be partitioned into 7 gates à 4 transitions. By analogy for $|B|=1$, we decompose the 20 transitions into 5 gates.}
\end{table}%

\noindent To complete the estimate for time evolving the two-flavor theory with trapped-ion qudits, we note on the side that the diagonal, local terms $\hat{\calN}_B$ and $\hat{\calN}_I$ in Eq.(\ref{eq:2flavor_partition_function}) can be treated as in the one-flavor case: If we decompose $\hat{\calN}_B = \sum_{x,i} |i\rangle\langle i|\otimes\omega_{x,i}$, where $\omega_{x, 0} = 0_{50\times50}$, $\omega_{x, 1} = -I_{20\times 20}\oplus I_{20\times 20} \oplus 0_{10\times 10} = -\sum_{i=0}^{19} \mathcal{Z}_{i,20+i}$ and $\omega_{x, 2} = -2\mathcal{Z}_{0,1}$, the gate decomposition in terms of diagonal single qudit rotations becomes visible, where a similar decomposition holds for $\hat{\calN}_I$.
  
 \subsubsection{Mapping to qumodes}
 \noindent In view of the 92-dimensional state space of the two-flavor theory (see Sect.\ref{sec:background}), mapping the theory to qumode degrees of freedom with a theoretically infinite dimensional Hilbert space appears promising. In fact, one could be tempted to simply generalize the relation Eq.(\ref{eq:h_meson_in_bosonic_ops}) obtained for $N_f=1$ by defining $\hat{J}_{Q_k,x}^{(i)+} = a^{\dagger (i,k)}_{1,x}\hat a^{(i,k)}_{2,x}$ to arrive at
\beq
\calH \sim \sum_{\langle x,y\rangle}\sum_{i,j}
  |i\rangle\langle i|_x |j\rangle\langle j|_y
 \sum_{Q_k} \lr{\hat  a^{\dagger (i,k)}_{1,x}\hat a^{(i,k)}_{2,x}\hat a^{\dagger (j,k)}_{2,y}\hat a^{(j,k)}_{1,y} + h.c.}.\nonumber
\label{eq:h_meson_nf2_in_bosonic_ops}
\eeq
However, this expression does not reproduce the correct matrix elements. Contrary to the one-flavor case, where the $J^{+/-}$ simply fulfilled the algebra of a spin-$N_c/2$ representation (c.f. Sect.\ref{sec:background}), the $\hat{J}_{Q_k,x}^{(i)+}$ in the $N_f = 2$-theory are reducible and decompose into product representations whose structure still has to be determined~\cite{WUproc2021, WUproc2022}. It is conceivable that the Jordan-Schwinger-map for the theory with $N_f=2$ requires several qumode operators $a_i^{k}$ per pion current $Q_k$ and -- following the discussion in Sect.\ref{subsec:nf1_qumodes} -- is likely to be expensive in terms of the number of gates which concludes our discussion for the two-flavor case.

%% file: Conclusion.tex
\section{Conclusion and Outlook}
\begin{table}
  \caption{\label{tab:storage_summary} Storage requirement for lattice volume $N$}
  \centering
  \begin{threeparttable}
 
    \begin{tabular}{c@{\qquad}c@{\qquad}c}
      Information Carrier & $N_f=1$ & $N_f=2$ \\ \midrule\midrule 
      Qubit & $3N$ & $(3+6)N$\tnote{*}\\
     \cmidrule(l r){1-3}
      Qudit $(d>2)$ & $N$ & $N+N$\tnote{**} \\ \cmidrule(l r){1-3}
      Qumode & $2N+N$\tnote{**} &  $>2N$\\ \midrule\midrule
    \end{tabular}

\begin{tablenotes}
  \item[*] We used unary encoding of the three ancilla states representing the baryon sectors $|B|$.
  \item[**] With mixed architecture, i.e. either mixed-qudit or qubit-qumode.
  \end{tablenotes}
  \end{threeparttable}
  \end{table}
  
  \begin{table}
  \caption{\label{tab:gc_summary}Entangling gate count for one mesonic $nn$-Trotter step}
  \centering
  \begin{threeparttable}
    \begin{tabular}{c@{\qquad}c@{\qquad}c}
      Information Carrier & $N_f=1$ & $N_f=2$ \\ \midrule\midrule 
      Qubit & $\mathcal{O}(10)$ &$\mathcal{O}(10^6)$\\
     \cmidrule(l r){1-3}
      Qudit $(d>2)$ & 2 & $\mathcal{O}(10^2)$ \\ \cmidrule(l r){1-3}
      Qumode & $\mathcal{O}(10^2)-\mathcal{O}(10^3)$\tnote{*} & --\\ \midrule\midrule
    \end{tabular}
  \begin{tablenotes}
  \item[*] Depending on the availability of the quartic gate.
  \end{tablenotes}
  \end{threeparttable}
  \end{table}
We summarize the relevant information on the resource requirement for the digitization of strong coupling lattice QCD with $N_f=1$ and $N_f=2$ flavors of staggered quarks in Table \ref{tab:storage_summary} and Table \ref{tab:gc_summary}. As qudits with $d>2$ offer the possibility to represent a multidimensional (local) state space, their resource requirements in terms of storage are favorable when compared to 2-level systems and qumodes. For the latter it should be noted that the scaling of course depends on the particular mapping chosen to encode the local, discrete state space of our models (angular momentum states) in the Fock-states of quantum harmonic oscillators, where we used the Jordan-Schwinger-map. For continuous variables, a different encoding may result in a more efficient scaling, as is the case for the non-linear sigma model~\cite{Jha2023a, Jha2023}. The entangling gate count in the qumode case (Table \ref{tab:gc_summary}) shows that the availability of non-Gaussian gates (in particular the quartic gate) has a high impact on the scaling. It remains to be seen if a measurement based approach as advocated in~\cite{Abel2024} offers a practical alternative. Conversely, for the qudit-case, it was the simultaneous driving of transitions via generalized MS-gates that allowed to reduce the gate count by at least one order of magnitude -- a reduction that is achieved by more complex experimental control. With qubits as information carrier, one main driver of gate count is the partitioning step for large Hamiltonians. Generally graph coloring is preferred which quickly becomes too expensive with the library we used~\cite{Hagberg2008} and we thus reverted to sequential approaches~\cite{Berg2020}. This partitioning or clustering is an active field of research and progress driven by the different optimization goals (e.g. simultaneous measurement or gate count reduction as in our case) will likely reduce the qubit result for $N_f=2$ in Table \ref{tab:gc_summary}.

As our study solely considers the resource requirements from an implementation point of view, an extension could  take into account the influence of errors on different platforms and existing error mitigation techniques. Another aspect that certainly deserves attention is the question of state preparation on different devices. In particular, for the photonic case, preparing states with different photon occupation number to encode the mesonic or baryonic sector, is an interesting subject of study.

%% file: Appendix.tex
\section{Digitization using qumodes}
\subsection{Exact gate decomposition using qumodes, gate count} 
\label{subsec:exact_decompo}
All of the eight terms appearing in Eq.(\ref{eq:h_meson_in_quadratures}) can be rewritten as $\hat x_{1,x} \hat x_{2,x} \hat x_{1,y} \hat x_{2,y}$ by means of unitary conjugation with Fourier gates, s.t. for example
\beq
e^{i\delta t \hat x_{1,x} \hat p_{2,x} \hat x_{1,y} \hat x_{2,y}} = F_{2,x}e^{i\delta t \hat x_{1,x} \hat x_{2,x} \hat x_{1,y} \hat x_{2,y}}F^\dagger_{2,x}\,.
\label{eq:hopping_term}
\eeq
Next we employ the identity
\beq
\hat x_{1,x} \hat x_{2,x} \hat x_{1,y} \hat x_{2,y} = \frac{1}{192}&[&(x_{1,x} + \hat x_{2,x} + \hat x_{1,y} + \hat x_{2,y} )^4 + (x_{1,x} + \hat x_{2,x} - \hat x_{1,y} - \hat x_{2,y} )^4\nonumber\\
&+&(x_{1,x} - \hat x_{2,x} + \hat x_{1,y} - \hat x_{2,y} )^4 + (x_{1,x} - \hat x_{2,x} - \hat x_{1,y} + \hat x_{2,y} )^4\nonumber\\
&-&(x_{1,x} - \hat x_{2,x} - \hat x_{1,y} - \hat x_{2,y} )^4 - (x_{1,x} + \hat x_{2,x} + \hat x_{1,y} - \hat x_{2,y} )^4\nonumber\\
&-&(x_{1,x} + \hat x_{2,x} - \hat x_{1,y} + \hat x_{2,y} )^4 - (x_{1,x} - \hat x_{2,x} + \hat x_{1,y} + \hat x_{2,y} )^4]\,.
\label{eq:complete_quartic_sum}
\eeq
Each term appearing on the right can be decomposed via unitary conjugation, s.t.
\beq
e^{i\delta t (\hat x_{1,x}+\hat x_{2,x}+\hat x_{1,y}+\hat x_{2,y})^4} =   e^{2i\hat p_{1,x}\hat x_{2,y}}e^{2i\hat p_{1,x}\hat x_{1,y}}e^{2i\hat p_{1,x}\hat x_{2,x}} e^{i\delta t \hat x^4_{1,x}}e^{-2i\hat p_{1,x}\hat x_{2,x}}e^{-2i\hat p_{1,x}\hat x_{1,y}}e^{-2i\hat p_{1,x}\hat x_{2,y}}\,,
\label{eq:quartic_sum}
\eeq
which can be seen by making repeated use of the Baker-Campbell-Hausdorff formula $e^A B e^{-A} = B+[A,B]+\frac{1}{2}[A,[A,B]]+\ldots$. We note that $e^{i\hat p_k\hat x_l} = F_k e^{i\hat x_k \hat x_l}F^\dagger_k$, where the operations on the $rhs$ are already part of the instruction set. The hopping term Eq.(\ref{eq:hopping_term}) is hence already decomposed into elementary operations, if the quartic gate $e^{i\delta t \hat x^4_{j}}$, appearing in Eq.(\ref{eq:quartic_sum}), is part of the instruction set. Otherwise the latter needs to decomposed by means of an ancilla qumode, $\hat x_j^4 = (\hat x_j^2+\hat x_a)^2-\hat x_a^2-2\hat x_a\hat x_j^2$, where we can further use that
\beq
e^{i\delta t ( \hat x_j^2+\hat x_a)^2} = e^{2i\hat p_{a}\hat x_j^2} e^{i\delta t \hat x_a^2}e^{-2i\hat p_{a}\hat x_j^2}
\label{eq:decomp_sum}
\eeq 
and employ the identity~\cite{Kalajdzievski2018}
\beq
e^{3i\alpha^2t\hat p_a\hat x_j^2} = e^{2i\alpha \hat x_j \hat x_a} e^{it\hat p_a^3}e^{-i\alpha \hat x_j \hat x_a}e^{-it\hat p_a^3}e^{-2i\alpha \hat x_j \hat x_a}  e^{it\hat  p_a^3}e^{i\alpha \hat x_j \hat x_a}e^{-it\hat p_a^3} e^{i\frac{3}{4}\alpha^3t\hat x_j^3}, \alpha \in \mathbb{R}\,.
\label{eq:decomp_sum_pxsq}
\eeq
A decomposition of the quartic gate $Q(\delta t) = e^{i\delta t \hat x^4_{j}}$ using the relations Eqs.(\ref{eq:decomp_sum}) and (\ref{eq:decomp_sum_pxsq}) turns out to be costly with a total of 55 elementary operations, of both, single qumode and entangling two-qumode type. This will ultimately drive the main cost of a mesonic time evolution. Consequently, in addition to the quartic gate $Q$, the decomposition shown in Eq.(\ref{eq:quartic_sum}) will take 6 $CX$-gates which will amount to 4 Fourier gates $F$ and 6 $CZ$-gates of our instruction set Eq.(\ref{eq:inst_set}). As each term appearing on the $rhs$ of Eq.(\ref{eq:complete_quartic_sum}) can be decomposed analogously, its exponentiation will lead to massive cancellations and combinations, resulting in a total count of $8 Q, 19 CZ$ and $18 F$ for the exponential of Eq.(\ref{eq:complete_quartic_sum}).\\
We now trotterize the exponential of the mesonic hopping term in Eq.(\ref{eq:h_meson_in_quadratures}) into the eight non-commuting terms of the form Eq.(\ref{eq:hopping_term}) with appropriate combinations of Fourier gates $p(F)$, i.e.
\beq
e^{i\delta t \sum_{\{i_k\}}(\hat  q^{i_1}_{1,x}\hat q^{i_2}_{2,x}\hat q^{i_3}_{2,y}\hat q^{i_4}_{1,y}) s_{\{i_1\ldots i_4\}}} = \prod_{\{i_k\}}p(F)e^{i\delta t \hat x_{1,x} \hat x_{2,x} \hat x_{1,y} \hat x_{2,y}  s_{\{i_1\ldots i_4\}}}p(F^\dagger) +\mathcal{O}(\delta t^2)\,.
\label{eq:meson_trotter_qumode}
\eeq
Given our considerations above, this can be done with cancellations in the appearing Fourier gates and CZ gates, s.t. the total gate count for the above operation is given by $64 Q, 147CZ, 149F$. If we consider the single qubit gate Q as one operation, then the corresponding gate depth is given by 348, as only interjacent Fourier gates can be carried out in parallel.